\documentclass[fleqn,usenatbib]{mnras}
\usepackage{mathptmx}
\usepackage{multirow}
\usepackage[T1]{fontenc}
\usepackage{soul}

\DeclareRobustCommand{\VAN}[3]{#2}
\let\VANthebibliography\thebibliography
\def\thebibliography{\DeclareRobustCommand{\VAN}[3]{##3}\VANthebibliography}

\usepackage{graphicx}
\usepackage{amsmath}
\usepackage{amssymb}

\title[GWs from GRB Magnetar Spindown]{Imprints of gravitational waves from magnetar spindown in GRB X-ray afterglows}

\author[S.J. Hou et al.]{
Shu-Jin Hou,$^{1}$\thanks{E-mail: houshujingrb@163.com}
Rui Xue,$^{2}$\thanks{E-mail: ruixue@zjnu.edu.cn}
Fang-Kun Peng,$^{3}$\thanks{E-mail: pengfk@ahnu.edu.cn}
Zhi-Gang Li,$^{1}$
Tong Liu,$^{4}$\thanks{E-mail: tongliu@xmu.edu.cn}
and Ren-Xin Xu$^{5,6}$\thanks{E-mail: r.x.xu@pku.edu.cn}
\\
$^{1}$Department of Physics and Electronic Engineering, Nanyang Normal University, Nanyang, Henan 473061, China\\
$^{2}$Department of Physics, Zhejiang Normal University, Jinhua 321004, China\\
$^{3}$Department of Physics, Anhui Normal University, Wuhu 241002, China\\
$^{4}$Department of Astronomy, Xiamen University, Xiamen, Fujian 361005, China\\
$^{5}$State Key Laboratory of Nuclear Physics and Technology, School of Physics, Peking University, Beijing 100871, China\\
$^{6}$Kavli Institute for Astronomy and Astrophysics, Peking University, Beijing 100871, China
}

\date{}

\pubyear{2026}

\begin{document}
\label{firstpage}
\pagerange{\pageref{firstpage}--\pageref{lastpage}}
\maketitle

\begin{abstract}
Given that newborn magnetars are considered potential central engines of gamma-ray bursts (GRBs), there is strong motivation to identify gravitational wave (GW) signatures within GRB samples. If the X-ray afterglow of a GRB is powered by a magnetar, and the initial spindown of the magnetar is dominated by the GW radiation induced by $r$-mode instability or magnetic-field-induced deformation, the decay of the X-ray flux would record the information of the GW radiation. We find that GRB 130603B potentially represents a rare and precious case where the spindown of the central magnetar is dominated in-turn by $r$-mode and magnetic distortion-induced GW radiation.  By fitting the X-ray light curve of GRB 130603B in this model, we obtain the initial spin period of magnetar $\sim 5.3\times 10^{-4}$ s, the effective dipole magnetic field strength $\sim 5.2\times 10^{14}$ G, the ellipticity of the magnetar $\sim 1.3\times 10^{-4}$, and the amplitude of $r$-mode oscillation $\sim3.3\times 10^{-2}$. It may serve as a reliable approach for investigating neutron star physics by comparing the parameters estimated using the method presented in this manuscript with those obtained from future GW observations.
\end{abstract}


\section{Introduction}\label{sec1}
Gravitational wave (GW) radiation generated by distorted neutron stars (NSs) has not been detected in the Milky Way till now \citep[e.g.,][]{2014ApJ...785..119A,2017PhRvD..96f2002A,2019ApJ...875..160A,2021ApJ...913L..27A}. This non-detection suggests that Galactic NSs have undergone effective spindown and maintain predominantly axisymmetric structures. As emphasized by \citet{2015PASA...32...34L}, rapid rotation and substantial quadrupole deformation constitute the two essential conditions for NSs to become detectable GW sources, making millisecond magnetars prime candidates. However, millisecond magnetars are rare. They are merely a brief period in the long life of NSs due to the effective spindown through, for example, magnetic dipole radiation. Without direct observation of GWs from NSs, other indirect sources are worth of considering, for example, gamma-ray bursts (GRBs).

The connection between magnetars and GRBs has been well established through theoretical studies \citep[e.g.,][]{1992Natur.357..472U,1992ApJ...392L...9D, 1998PhRvL..81.4301D,1998A&A...333L..87D,2001ApJ...552L..35Z,2007MNRAS.380.1541B,2011MNRAS.413.2031M}. Magnetars, with the adequate rotational energy ($E_{\rm rot} \sim 10^{52}$ erg) and the capacity to power magnetic dipole luminosity as high as $\sim 10^{49}\rm erg \ s^{-1}$, attract widespread attentions in explaining the X-ray afterglow emission of GRBs. Particularly, the plateau segments of GRB X-ray light curves could be powered by the spindown winds of nascent magnetars \citep[e.g.,][]{2010MNRAS.409..531R,2010ApJ...715..477Y,2020ApJ...901...75D,2021ApJ...922..102H}.

Electromagnetic (EM) radiation may be not the only effect way that makes a nascent millisecond magnetar slowing down. GW radiation originating from distinct but coupled physical processes may also work \citep[e.g.,][]{1998ApJ...502..708A,2009ApJ...702.1171C,2013ApJ...771...86G,2013PhRvD..88f7304F,2017ApJ...843L...1L,2016PhRvD..93d4065G,2017ApJ...851L..16A,2018PhRvL.120z1103M}. The $r$-mode instabilities and magnetic-field-induced deformations are two potential factors, which can induce intense GW radiation. Since these two processes depend critically on the spin periods of magnetars, the corresponding GW emission can significantly modify the spindown trajectory, thereby imprinting detectable signatures on the X-ray afterglow light curves. Therefore, the X-ray afterglows of GRBs serve as promising tools for indirectly studying GW emission from NSs \citep[e.g.,][]{2014ApJ...785...74L,2017ApJ...835..181L,2022A&A...666A.138L,2025arXiv250711110L}.

\citet{2016MNRAS.463..489H} and \citet{2022MNRAS.513.1365X} studied GRB light curves within the magnetar model through the following five processes, the EM radiation, the GW radiation from $r$-mode, the GWs from magnetic distortion, the EM radiation + the GW radiation from $r$-mode, and the EM radiation + the GW radiation from magnetic distortion. The corresponding evolutions of X-ray afterglow light curves are as follows: $L \propto t^{0} \rightarrow t^{-2}$, $L \propto t^{0} \rightarrow t^{-2/3}$, $L \propto t^{0} \rightarrow t^{-1}$, $L \propto t^{0} \rightarrow t^{-2/3} \rightarrow t^{-2}$, and $L \propto t^{0} \rightarrow t^{-1} \rightarrow t^{-2}$. However, no case containing the GW radiation both induced by $r$-mode instability and magnetic distortion, has been found till now. While this scenario likely provides a more realistic representation of a magnetar, such sources are expected to be exceptionally rare in astrophysical populations.

Alternatively, the braking index $n$ \citep{1985Natur.313..374M, 2017ApJ...843L...1L, 2018PhRvD..98d3011S} can also be invoked to indicate which mechanism dominates the NS spindown. The braking indices, corresponding to the energy loss caused by magnetic dipole radiation, GW radiation induced by magnetic deformation, and $r$-mode GW radiation, are \textit{n}=3, \textit{n}=5 and \textit{n}=7, respectively \citep{2023LRR....26....3R, 2025PhRvD.112b3020B}. Based on the fiducial millisecond magnetar model, \citet{2017ApJ...843L...1L} measured the braking index $n=3 $ for GRB 130603B, and \citet{ 2019ApJ...871...54L} found that the braking index was not a constant and satisfied a normal distribution ($n_{c}$ = 4) based on a sample of short GRBs. \citet{2019ApJ...871..160L} measured the braking index $n=7$ for GRB 090510 and \citet{2018PhRvD..98d3011S} used the data of GRB 140903A to give the upper limit of the ellipticity.

In this \emph{Letter}, by scrutinizing the X-ray light curves in \textsl{Swift} GRB sampe, we found a rare and previous case GRB 130603B, whose emission possibly was shaped simultaneously by the GW radiations induced by $r$-mode instability and magnetic distortion. In section \ref{sec2}, we present the basic mathematic formalism and theoretical framework under the GRB-magnetar scenario. In section \ref{sec3}, we constrain the physical parameters of gravitational wave emission with GRB 130603B. A brief summary is given in section \ref{sec4}.

\section{Electromagnetic radiation and gravitational wave radiation}\label{sec2}

After massive star collapse or the merger of binary NSs, a rapidly rotating magnetar might form at the center \citep[e.g.,][]{2011PhRvD..83d4014G, 2012LRR....15....8F,2021ApJ...908..106L,2023ApJ...952..156S}. The nascent magnetar loses its rotational energy through multiple channels, including EM radiation and GW emission. There may be a variety of causes of GW radiation from magnetars, among which, $r$-mode oscillation and magnetic distortion are usually considered to be two main ways to induce the GW emissions.

The GW radiation induced by magnetic distortion is approximately expressed as the following equation, during the period when the magnetic field has not yet undergone significant evolution, which is given by \citep[e.g.,][]{1963PhRv..131..435P,M08}
\begin{eqnarray}\label{2}
L_{\rm gw,\epsilon}=\frac{32GI^{2}\epsilon^{2}\Omega^{6}}{5c^{5}},
\end{eqnarray}
where $G$ is the gravity constant, $I$ is the moment of inertia, $\Omega$ is the angular velocity, $\epsilon$ is the magnetically induced ellipticity, and $c$ is the speed of light.

The GW radiation from the leading mode of $r$-mode oscillation \citep[e.g.,][]{1998PhRvD..58h4020O,2016MNRAS.463..489H}, reads
\begin{eqnarray}\label{3}
L_{\rm gw,r}\approx \frac{96\pi}{15^{2}}\left ( \frac{4}{3} \right )^{6} \frac{GMR^{4}\tilde{J}^{2}I}{c^{7}\tilde{I}}a^{2}\Omega^{8},
\end{eqnarray}
where $M$ is the NS mass, $\tilde{J}=\frac{1}{MR^{4}}\int_{0}^{R}\rho r^{6}dr$ with $\rho$ being the mass density, $\tilde{I}=\frac{8\pi}{3MR^{2}}\int_{0}^{R}\rho r^{4}dr$, $a$ is the amplitude of the oscillation, and $R$ is the equatorial radius.

The power of magnetic dipole radiation is
\begin{eqnarray}\label{4}
L_{\rm em}=\frac{B_{\rm eff}^{2}R^{6}\Omega^{4}}{6c^{3}},
\end{eqnarray}
where $B_{\rm eff}$ is the effective dipole magnetic field strength on the NS polar cap \citep[e.g.,][]{2001ApJ...552L..35Z,2014MNRAS.439.3916M}. When the spindown of the magnetar is dominated by the magnetic dipole radiation only, the EM radiation luminosity evolves as
\begin{eqnarray}\label{5}
L_{\rm em}(t)=L_{\rm em,0}\left(1+\frac{t}{\tau_{\rm em}}\right)^{-2},
\end{eqnarray}
where $L_{\rm em,0}=\frac{\eta I\Omega_0^2}{2\tau_{\rm em}}$ with $\tau_{\rm em}=\frac{E_{\rm rot}}{L_{\rm em,0}}=\frac{3c^3I}{B_{\rm p}^2R^6\Omega_0^2}$, and $\Omega_0$, are the luminosity and angular frequency at the initial time $t=0$.

Combining the total energy release determined by equation (\ref{2})-(\ref{4}),
the evolution of the NS spindown should be given by
\begin{eqnarray}\label{6}
L_{\rm em}+L_{\rm gw,\epsilon}+L_{\rm gw,r}=-I\Omega\frac{d\Omega}{dt},
\end{eqnarray}
where $\frac{d\Omega}{dt}$ are the spin velocity time derivative \citep[e.g.,][]{1983bhwd.book.....S,2001ApJ...552L..35Z}.
Equation (\ref{6})  can also be rewrote as
\begin{eqnarray}\label{7}
\frac{d\Omega }{dt}=\alpha \Omega^{3} +\beta \Omega^{5}+\gamma \Omega^{7},
\end{eqnarray}
where
\begin{eqnarray}\label{8}
\alpha=-\frac{B_{\rm eff}^{2}R^{6}}{6c^{3}I},
\end{eqnarray}
\begin{eqnarray}\label{9}
\beta=-\frac{32GI\epsilon^{2}}{5c^{5}},
\end{eqnarray}
and
\begin{eqnarray}\label{10}
\gamma=-\frac{96\pi}{15^{2}}\left ( \frac{4}{3} \right )^{6}
\frac{GMR^{4}\tilde{J}^{2}}{c^{7}\tilde{I}}a^{2}.
\end{eqnarray}

\begin{figure}
\centering
\includegraphics[width=0.5\textwidth]{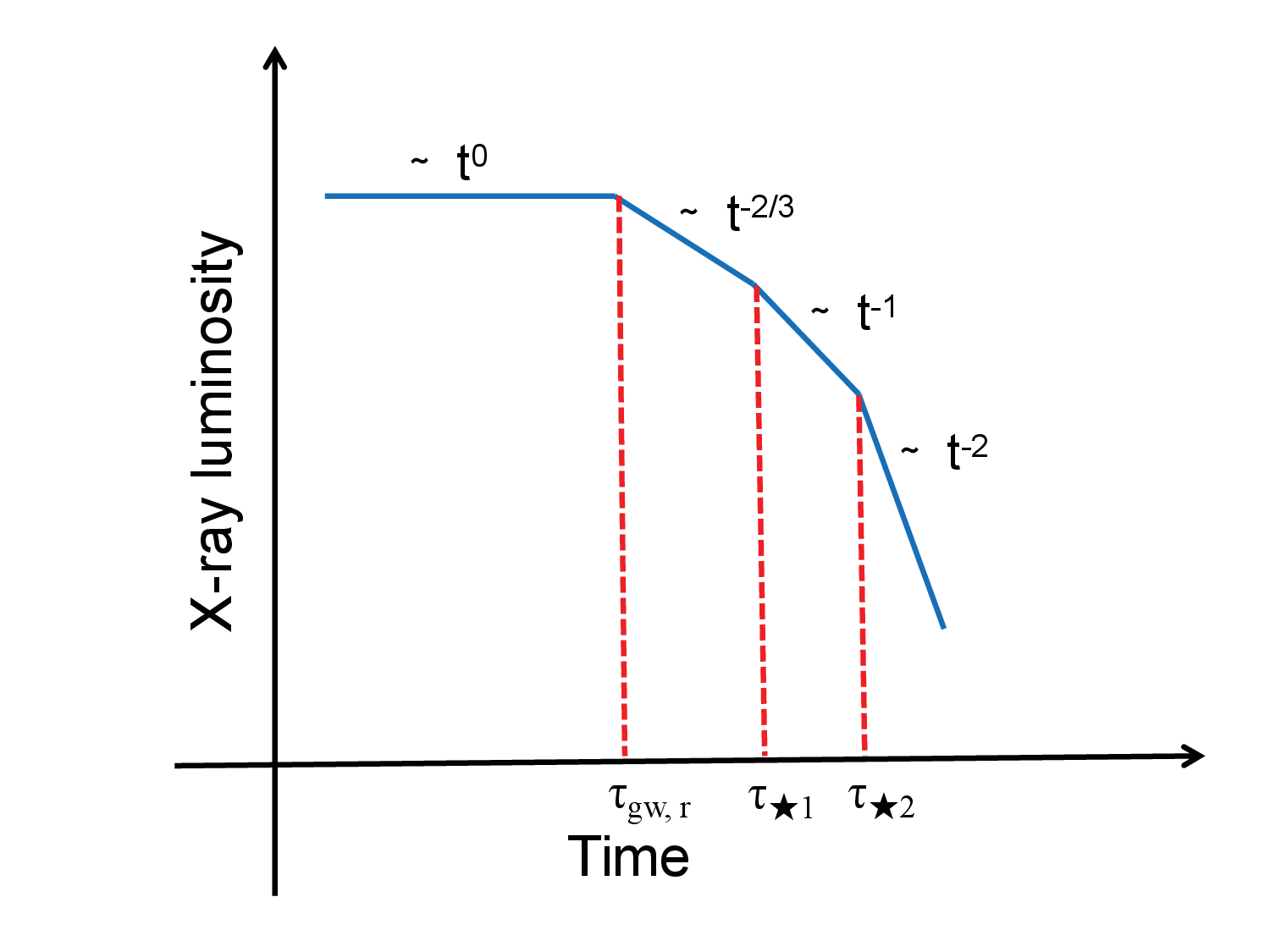}
\caption{Schematic diagram of the X-ray afterglow of GRB contributed by magnetar spindown. The first two parts are dominated by $r$-mode oscillation, the third part is dominated by magnetic distortion process, and the fourth part is dominated by magnetic dipole radiation. Here $\tau_{\rm gw,r}$ is the GW spindown timescale. The time $\tau_{\rm \star1}$  marks the transition where the spindown luminosity is no longer dominated by $r$-mode GWs but by GWs from magnetic distortion, while $\tau_{\rm \star2}$ marks the subsequent transition to dominance by magnetic dipole radiation.}
\label{fig.1}
\end{figure}

Based on the above equations, the different correlations between the luminosities and angular velocity like $L_{\rm em}\propto \Omega^{4}$, $L_{\rm gw,\varepsilon}\propto \Omega^{6}$, and $L_{\rm gw,r}\propto \Omega^{8}$ \citep{2016MNRAS.458.1660L}, imply that, in the early stages of magnetar evolution, $L_{\rm em} \ll L_{\rm gw,\varepsilon} \ll L_{\rm gw,r}$ might be satisfied. Since the angular velocity slows down with the time, these correlations mean that the decay of $L_{\rm gw,r}$ is fastest, the decay of $L_{\rm gw,\varepsilon}$ is slower, and the decay of $L_{\rm em}$ is the slowest. That is to say, the initial spindown is successively dominated by three mechanisms: first through $r$-mode induced gravitational-wave radiation, then by gravitational-wave radiation from magnetic deformation, and ultimately by electromagnetic radiation. If the spindown is governed by magnetic dipole radiation initially, the GW radiation will never dominate the spindown.

The existence of gravitational wave radiation during the plateau phase in GRB X-ray afterglow is widely considered. A key requirement for detecting this signature in a GRB X-ray light curve is that GW radiation must dominate the initial spindown evolution. This dominance ensures that the effect of GWs is recorded in the electromagnetic emission, thereby allowing GW signals to be confirmed indirectly through observed EM radiation observation.

In our model, $r$-mode oscillation dominates the evolution of spindown process at early times (the first stage), then the basic equations of evolution can be simplified as
 \begin{eqnarray}\label{11}
\left\{\begin{matrix}
L_{\rm gw,r}\approx -I\Omega\frac{d\Omega}{dt}, \\
L_{\rm em}=\frac{B_{\rm eff}^{2}R^{6}\Omega^{4}}{6c^{3}}.
\end{matrix}\right.
\end{eqnarray}
By solving Equation (\ref{11}), we can get the evolutionary equation of $L_{\rm em}$, which can be expressed as
\begin{eqnarray}\label{12}
	L_{\rm em}(t)=L_{\rm em,0}\left(1+\frac{t}{\tau_{\rm gw,r}}\right)^{-2/3},
\end{eqnarray}
where $\tau_{\rm gw,r}=\frac{E_{\rm rot}}{L_{\rm gw,r}}$ is the GW spindown timescale. Before $\tau_{\rm gw,r}$, there is $L_{\rm em}\propto t^{0}$ and it is a plateau. After the plateau, there is a break, which is $L_{\rm em}\propto t^{-2/3}$.

In this case, there is a transition timescale, $\tau_{\rm \star1}$, which corresponds to the spindown luminosity shifting its primary source from $r$-mode-generated GWs to GWs generated by magnetic distortion. At time $\tau_{\rm \star1}$, the emission satisfies $L_{\rm gw,r}(\tau_{\rm \star1})=L_{\rm gw,\varepsilon}(\tau_{\rm \star1})$. From $\tau_{\rm gw,r}$ to $\tau_{\rm \star1}$, there is $L_{\rm em}\propto t^{-2/3}$.

After the domination of the $r$-mode GWs emission, the GWs from magnetic distortion process dominates, then the basic equations of evolution can be written as
\begin{eqnarray}\label{13}
\left\{\begin{matrix}
L_{\rm gw,\epsilon}\approx -I\Omega\frac{d\Omega}{dt}, \\
L_{\rm em}=\frac{B_{\rm eff}^{2}R^{6}\Omega^{4}}{6c^{3}}.
\end{matrix}\right.
\end{eqnarray}
There also exists a transition timescale $\tau_{\rm \star2}$, which is the time of the spindown luminosity changes from these magnetically-distorted GWs to magnetic dipole radiation. At $\tau_{\rm \star2}$, the following equation holds $L_{\rm em}(\tau_{\rm \star2})=L_{\rm gw,\varepsilon}(\tau_{\rm \star2})$. From $\tau_{\rm \star1}$ to $\tau_{\rm \star2}$, the light curve decays as $L_{\rm em}\propto t^{-1}$. After the $\tau_{\rm \star2}$, the spindown is dominated by magnetic dipole radiation, the light curve goes like $L_{\rm em}\propto t^{-2}$.

In our model, the X-ray light curves go through three stages, and the decay indices have four phases, as shown in Figure \ref{fig.1}. In the first two parts of the light curve, $r$-mode oscillation is dominant, with corresponding evolution indices of $\sim0$ and $\sim-2/3$. The third part is dominated by magnetic distortion process and the fourth part is dominated by magnetic dipole radiation, with corresponding evolution indices of $\sim-1$ and $\sim-2$, respectively. Luckily, it is found that GRB 130603B is the rare and previous sample exhibiting an evolution pattern of $\sim 0\rightarrow \sim -2/3\rightarrow \sim -1\rightarrow\sim -2$ (see Figure \ref{fig.2}).

The direct GW observations offer the most powerful way to test our models. The GW strain from spinning magnetar with magnetic distortion \citep[e.g.,][]{1998PhRvD..58h4020O, 2009ApJ...702.1171C} is
\begin{eqnarray}\label{14}
h\textsuperscript{$\epsilon$}(t)=\frac{4G\Omega(t)^2}{c^4d}I\epsilon,
\label{eq:ht}
\end{eqnarray}
where $d$ is the distance from the source to the Earth. For long-lived GW transient sources, this amplitude can be approximately expressed as \citep[e.g.,][]{1998PhRvD..58h4020O, 2009ApJ...702.1171C}
\begin{eqnarray}\label{15}
&&h\textsuperscript{$\epsilon$} \approx 4.18 \times 10^{-25} \\ \nonumber &&\times \left(\frac{\epsilon}{10^{-4}}\right) \left(\frac{f}{1\,\text{kHz}}\right)^2 \left(\frac{I}{10^{45}\,\text{g}\cdot\text{cm}^2}\right) \left(\frac{1\,\text{Mpc}}{d}\right),
\label{eq:h1}
\end{eqnarray}
where $f$ is the frequency of the GW signal. The GW strain with $r$-mode is
\begin{eqnarray}\label{16}
h\textsuperscript{r}(t)= \frac{32\sqrt{30\pi}}{75} \cdot \frac{GM}{c^5} \cdot \frac{\alpha R^3 \tilde{J} \Omega(t)^3}{d},
\end{eqnarray}
which is well approximated as
\begin{eqnarray}\label{17}
&&h\textsuperscript{r}\approx 7 \times 10^{-26} \\ \nonumber &&\times \left(\frac{a}{0.01}\right) \left(\frac{f}{\text{1 kHz}}\right)^3 \left(\frac{R}{10\,\text{km}}\right)^4 \left(\frac{1\,\text{Mpc}}{d}\right).
\label{eq:h2}
\end{eqnarray}
Therefore, the detection of the GW radiation from a sample like GRB 130603B will provide a direct test to our model if $\epsilon$,
$a$ and $\Omega$ can be estimated.

\section{Case study of GRB 130603B}\label{sec3}

\begin{figure*}
\centering
\includegraphics[width=0.50\textwidth]{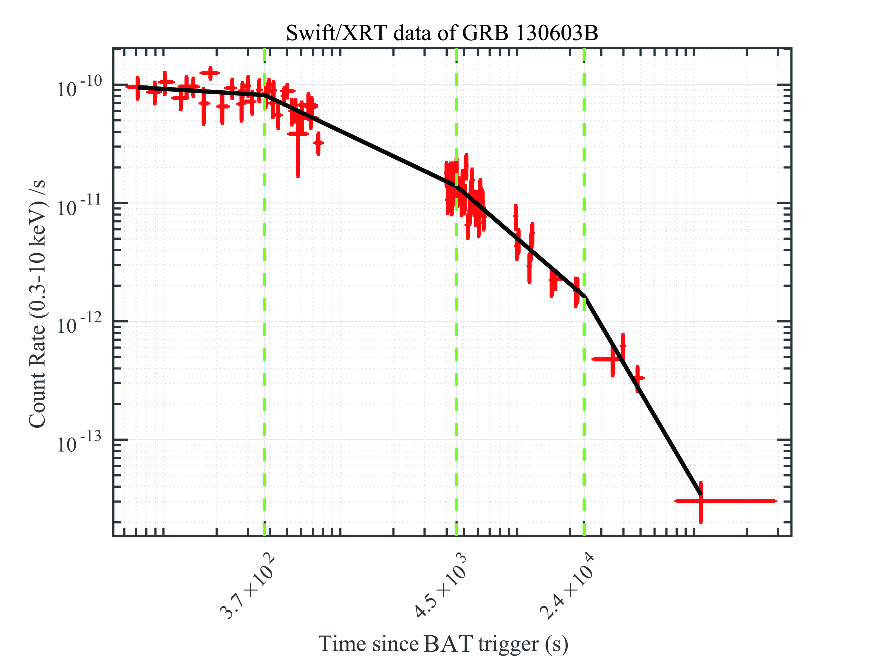}
\includegraphics[width=0.45\textwidth]{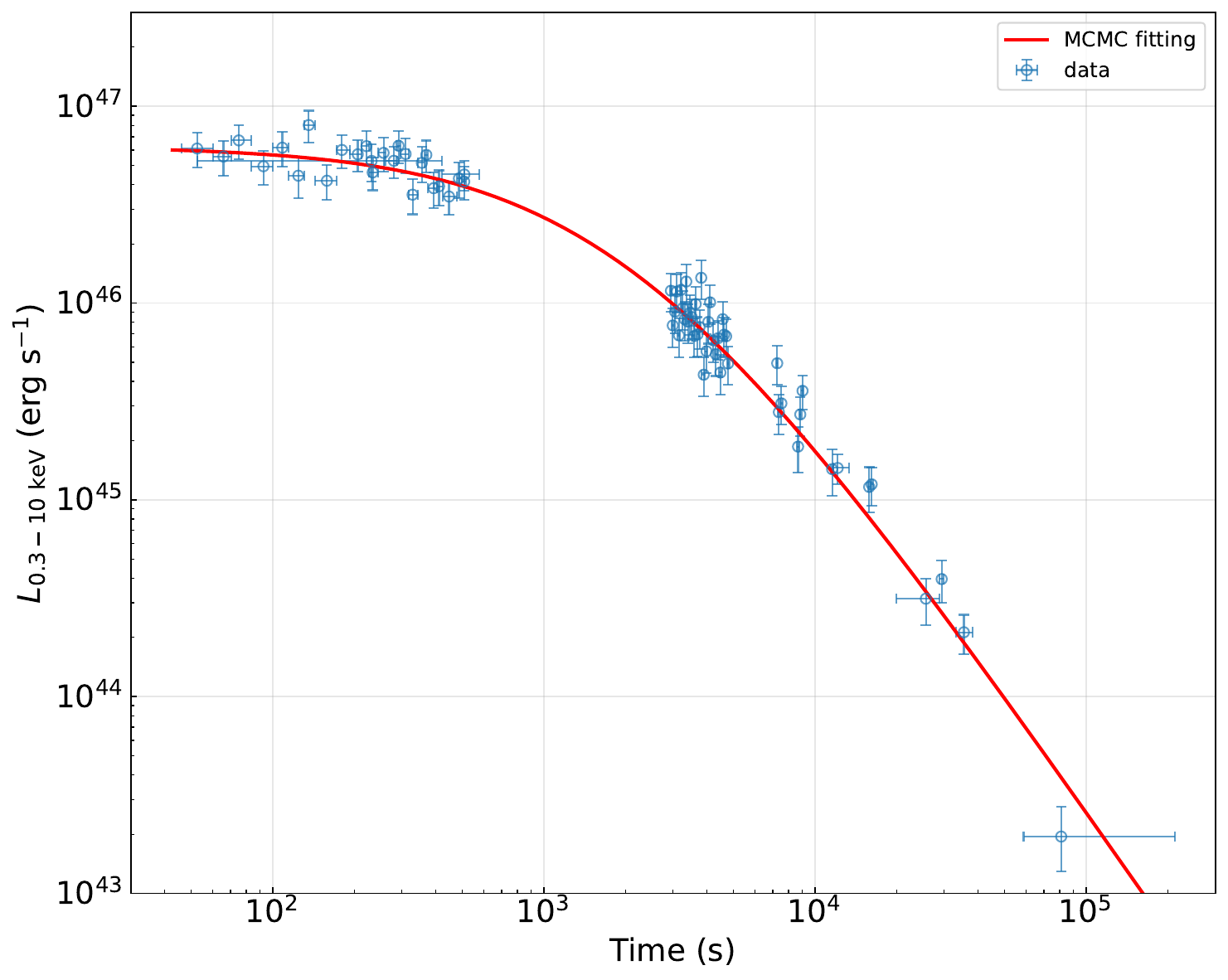}
\caption{Left panel: Fitting of the X-ray light curve of GRB 130603B with multi power-law functions. The green dashed lines indicate the corresponding break times. The evolution indices of the four segments are: $\sim 0.09, \sim 0.71, \sim 1.23, \sim 2.50$, respectively. Right panel: MCMC fitting on the unabsorbed luminosity of GRB 130603B with magnetar model based on the Equation (\ref{7}).}
\label{fig.2}
\end{figure*}

The X-ray plateau emission of GRBs afterglow usually may be from magnetar spindown due to energy release via EM radiation \citep{2001ApJ...552L..35Z}, although it may also be from the fall-back accretion onto black hole \citep{2013ApJ...767L..36W}. The diversity of energy sources results in the complexity of the GRB X-ray afterglow. Previous studies indicate that the X-ray emission in most GRBs is governed either purely by magnetic dipole radiation or by a combination of magnetic dipole radiation with one of the gravitational-wave mechanisms mentioned previously \citep[e.g.,][]{2014ApJ...785...74L,2016MNRAS.463..489H,2016MNRAS.458.1660L,2022MNRAS.513.1365X}. Fortunately, we discover a rare and precious sample, GRB 130603B. The sample is a short GRB with duration being $T_{90}\approx 0.18~\rm {s}$ \citep{2013GCN.14735....1M,2013Natur.500..547T} and redshift $z=0.356$ \citep{2013GCN.14747....1S}. The X-ray afterglow emission could be best-fitted by a four-segment power-law function with indexes comparable to those in Figure \ref{fig.1}. The best-fitting line with a four-segment power-law function was displayed in the left panel of Figure \ref{fig.2}, which shows a close agreement between the data points and the model lines (reduced $\chi^2 \simeq 1.47$). The best-fitted indices and their corresponding break times are listed in Table \ref{ta1}. The evolution of count rate over time was characterized by $\propto t^{-0.09} \rightarrow t^{-0.71} \rightarrow t^{-1.23} \rightarrow t^{-2.5}$, aligns well with indexes predicted by our theoretical framework in the error bars. We therefore identify GRB 130603B as a promising candidate source for our model.

\begin{table*}
\centering
\caption{Results of X-ray light curve of GRB 130603B with multi power-law functions fitting. The three break times, marked as green dashed lines in Fig. \ref{fig.2}, luminosities at corresponding break times, and four indexes are shown in this table.}\label{ta1}
\begin{tabular}{cccc}
\hline
 \ & Time ~(error) & Index ~(error) & Luminosity ~(error) \\
  \ & s &  & $10^{45}$ erg $\rm s^{-1}$ \\
\hline
$t_{\rm b,1}$ & 373 (300) & 0.09 (0.03) & 35.7 (2.65)\\
$t_{\rm b,2}$  & 4550 (3730)& 0.71 (0.07) & 6.07 (1.12)\\
$t_{\rm b,3}$  &$\rm 2.41 (1.06)\times 10^4$ & 1.23 (0.14) &  0.71 (0.21)\\
 &  & 2.50 (1.90) & \\
\hline
\end{tabular}
\end{table*}

\begin{figure*}
\centering
\includegraphics[width=0.85\textwidth]{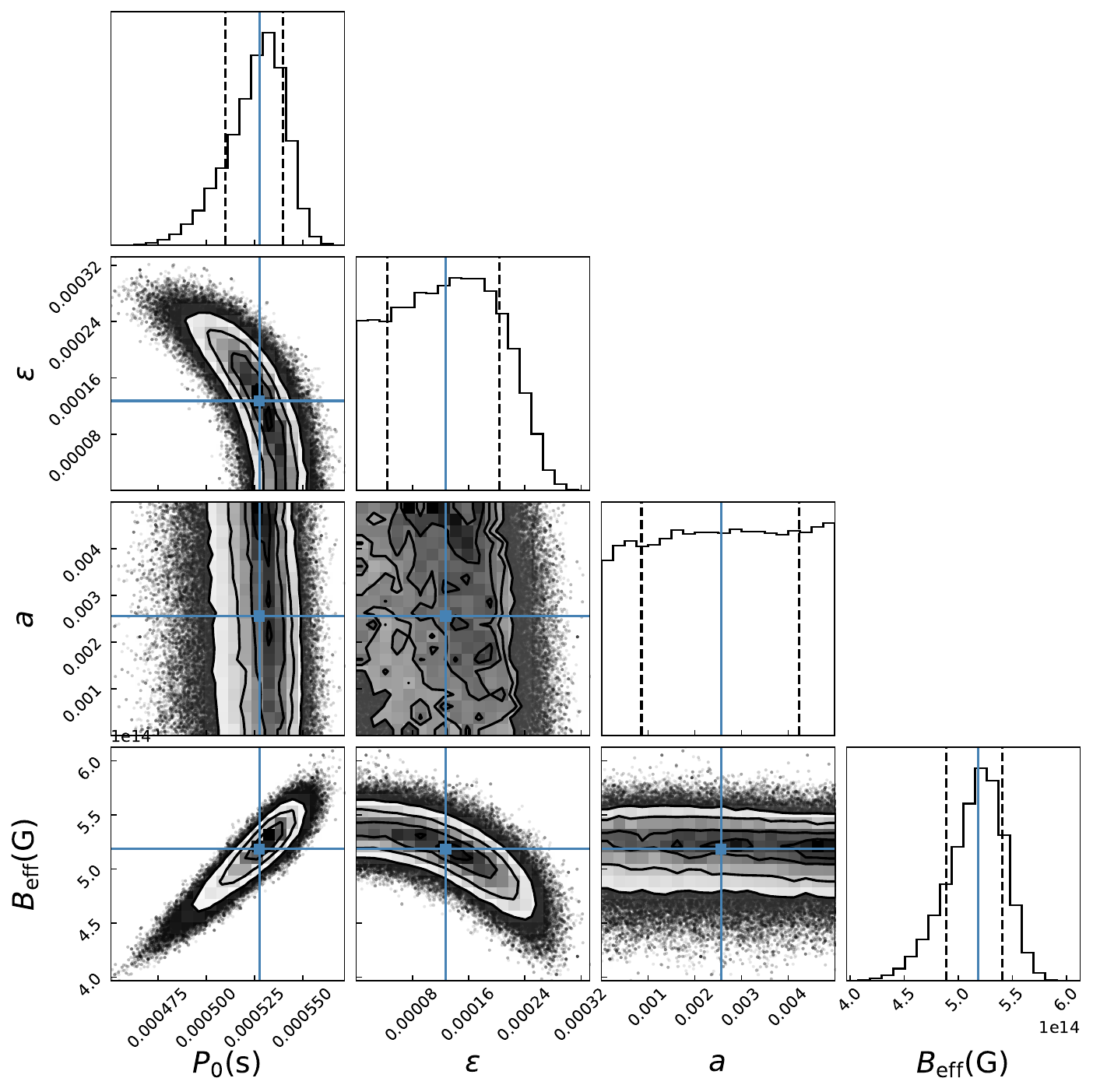}
\caption{Corner plot of the posterior probability distributions of the magnetar parameters: the initial period $P_0$, the ellipticity $\epsilon$, the amplitude of the oscillation from $r$-mode $a$, and the effective magnetic field $B_{\rm eff}$.}
\label{fig.3}
\end{figure*}

\subsection{MCMC fitting}

We first numerically solve Equation (\ref{7}) to model the magnetar's period evolution over time. Based on the framework of magnetic dipole radiation, we then fit the X-ray data of GRB 130603B to determine the underlying physical parameters of the magnetar. When applying our model to this sample, we find it contains too many free parameters. Therefore, some parameters must be fixed a priori under the physical requirements of magnetar. Considering the rarity of this sample, if the central engine of GRB 130603B is truly a magnetar, then this magnetar must be an extreme case within the magnetar population. We speculate that this magnetar should have a rotational period extraordinary close to the Kepler limit \citep{1992PhRvD..46.4161G} to invoke GW emission via $r$-mode ($L_{\rm gw,r}\propto \Omega^{8}$).

As indicated by the observation of GRB 170817A \citep{2017ApJ...848L..13A},
the short GRB 130603B should originate from a binary neutron star merger. The mass of the central object of GRB 130603B should be $\sim 2-3 M_{\odot}$. Since the magnetar-powered X-ray emission from GRB 130603B persists for $\sim 10^{5}\rm \ s$, our model requires the central magnetar¡¯s lifetime to be at least of the same order. Such a long-lived magnetar would imply its mass should be a little larger than the upper rest-mass limit for a neutron star. Based on the observations \citep{2010Natur.467.1081D, 2013Sci...340..448A, 2020NatAs...4...72C} and theoretical researches \citep{2018PhRvL.120q2703A}, the following parameters of the magnetar are selected
\footnote{We also examined alternative sets of magnetar parameter values, such as different mass and raduis of magnetar, and found that the changes in the best-fitted parameters are minor.}:  the mass $M=2.4{\rm M_{\odot}}$, the rotational inertia
$I\sim 3\times10^{45} {\rm g\;cm^{2}}$, and the equatorial radius $R \sim 13 {\rm km}$ \citep{2015PhRvD..92b3007C}.

We follow the results shown in \citet{1998PhRvD..58h4020O} and \citet{2016MNRAS.463..489H}, and adopt $\tilde{J}=0.02$ and $\tilde{I}= 0.3$. By taking these values, we employ the Markov Chain Monte Carlo (MCMC) technique using the publicly available Python package \textsl{emcee} \citep{2013PASP..125..306F} to constrain the rest magnetar parameters. The prior of parameters is uniformly distributed over a large range. The prior setting and best-fitting results should be under the requirements of magnetar physics as well. For example, the saturation amplitude $a$ is at a level of $10^{-4}-10^{-3} $\citep{2003ApJ...591.1129A,2007PhRvD..76f4019B,2009PhRvD..79j4003B}, and the correlation between ellipticity and magnetic field should be satisfied \citep{2002PhRvD..66h4025C}. Another key parameter is the dipole radiation efficiency $\eta$, which is treated as a fixed input rather than a free parameter in the fitting process. Referring to the earlier discussions and following the content in \citet{2019ApJ...878...62X}, we assume a constant X-ray (0.3¨C10 keV) efficiency of $\eta = 6 \times 10^{-4}$ for magnetic dipole radiation throughout this study. Specifically, we set $P_0 \in [10^{-1},30]~\rm ms$, $\epsilon \in [1.0\times 10^{-7}, 1.0\times 10^{-2}]$, $a \in [1.0\times 10^{-6}, 5.0\times 10^{-3}]$, and $ B_{\rm eff} \in [1.0\times 10^{13}, 1.0\times  10^{16}]~\rm G$. The results of the MCMC fitting are presented in the right panel of Figure \ref{fig.2}. The model shows excellent agreement with the observational data, yielding a reduced chi-square value of $\chi^2_{\rm red} \simeq 1.11$. The posterior probability distributions of the magnetar parameters are presented in the corner plot of Figure \ref{fig.3}. Three parameters, $P_0$, $\epsilon$, $B_{\rm eff}$, are constrained. {We get the initial spin period of the magneter $\sim 5.3\times 10^{-4}$ s, the effective dipole magnetic field strength $\sim 5.2\times 10^{14}$ G, and the ellipticity of the magnetar $\sim 1.3\times 10^{-4}$. In contrast, the amplitude $a$ of the $r$-mode is poorly constrained.
We note that, at the first break of the X-ray light curve, there is
\begin{eqnarray}\label{aa}
L_{\rm gw,r}(\Omega_{0})\cdot\tau_{\rm gw,r}=\frac{1}{2}I\Omega_{0}^{2}.
\label{eq:haa}
\end{eqnarray}
Substituting equation (\ref{3}) into equation (\ref{aa}), using the estimated value of $\tau_{\rm gw,r}$ shown in table \ref{ta1} and the best-fitting initial spin period $P_0$, we get $a\sim 3.3\times 10^{-2}$.

\subsection{Discussion: our estimations versus theoretical expectations}
Compared with theoretical expectations, the two gravitational wave radiation parameters estimated in this work: $\epsilon\sim 1.3\times 10^{-4}$ and $a\sim 3.3\times 10^{-2}$, appear to be relatively large. If the magnetar model is indeed applicable, the scarcity of events like GRB 130603B would imply that the central engine must be a special magnetar with extreme parameters. In that case, the seemingly large values obtained here may be justified, as detailed in the following discussion.

(i) The magnetic distortion of a magnetar may be mainly contributed by the magnetar internal toroidal field \citep{2002PhRvD..66h4025C}, e.g., $\epsilon\sim 1.6\times 10^{-6}<B_{\rm t}^{2}>/(10^{30}\rm G^{2})$. If the internal field strength is comparable to the external field estimated here ($\sim 5.2\times 10^{14}\rm G$), the theoretical ellipticity would indeed be far below $1.3\times 10^{-4}$. However, the surface magnetic field of a magnetar could be significantly weaker than its internal toroidal component \citep{1995MNRAS.275..255T,1996ApJ...473..322T}. If the internal field is in fact tens of times stronger than the surface field, as suggested in recent studies (e.g., \citet{2022A&A...666A.138L}), the tension on ellipticity between the theoretical prediction and the observational constraint would be substantially alleviated.

 ii) As emphasized above, the scarcity of the sample like GRB 130603B implies that the magnetar parameters obtained through fitting may be relatively extreme. When the spin period is close to the Kepler limit ($\sim 0.5 \ {\rm ms}$; e.g., see \citealt{1992PhRvD..46.4161G}), the $r$-mode amplitude estimated here is compatible with both the constraint $a<0.025 (B_{\rm eff}/10^{14}\ {\rm G})(P_{0}/1\ {\rm ms})^{2}$ from \citet{2016MNRAS.463..489H}, and the theoretical prediction $a\approx 8\times 10^{-3}(P_{0}/1\ {\rm ms})^{-5/2}$ derived from second-order perturbation theory, which includes the nonlinear coupling of short-wavelength inertial modes \citep{2003ApJ...591.1129A}. Besides, such extreme parameters may also serve as possible signs for the existence of non-traditional processes in magnetars. The oscillation amplitude could be magnified by additional physical mechanisms, possibly including magnetic-field-driven channels (e.g., \citealt{2000ApJ...543..386H}) or unconventional non-parametric coupling (e.g., \citealt{2001PhRvL..86.1152L}).

\section{Summary}\label{sec4}

In this \emph{Letter}, we take the X-ray afterglow of GRB 130603B as a case study to investigate the GW and EM emissions from a newborn magnetar within the GRB-magnetar framework. We model an evolutionary sequence wherein the dominant energy-loss mechanism transitions from GW radiation (driven by $r$-mode instability) to GW radiation (from magnetic distortion), and ultimately to magnetic dipole radiation. Within this model, the magnetar-powered X-ray light curve naturally divides into four distinct phases characterized by a temporal decay index that evolves as $\sim 0 \rightarrow \sim -2/3 \rightarrow \sim -1 \rightarrow \sim -2$.

The short burst GRB 130603B is hypothesized to harbor a magnetar as central engine \citep{2013Natur.500..547T, 2017ApJ...843L...1L}. Its X-ray afterglow is characterized by four distinct power-law phases, with the luminosity evolving as $L \propto t^{-0.09} \rightarrow t^{-0.71} \rightarrow t^{-1.23} \rightarrow t^{-2.5}$.
This light-curve behavior matches our model, providing indirect evidence for an early spindown dominated by GW radiation (first from r-mode instability, then from magnetic distortion) and a later transition to EM-dominated spindown.

We perform the MCMC fitting on the X-ray observation of GRB 130603B in the framework of GRB-magnetar scenario. The magnetar parameters are obtained: the initial spin period $\sim 5.3\times 10^{-4}$ s, the effective dipole magnetic field strength $\sim5.2\times 10^{14}$ G, the ellipticity $\sim 1.4\times 10^{-4}$, and the $r$-mode oscillation amplitude $\sim 3.3\times 10^{-2}$. The estimated $r$-mode amplitude and the ellipticity are extremely high. These results may occupy the most extreme region of the magnetar parameter space, which is consistent with the rarity argument within \textsl{Swift} GRB sample.

We note that our constraints are premised on two assumptions: the general applicability of the magnetar scenario to GRB plateaus and its specific relevance to GRB 130603B. The derived parameters therefore depend on these model premises. Nonetheless, this investigation remains valuable as it addresses the challenge of directly detecting GWs from newborn magnetars. The estimated GW strains ($h^{\epsilon} \sim 10^{-25}$ and $h^r \sim 10^{-26}$ at 1 Mpc) remain below current LIGO/Virgo detection thresholds for extragalactic sources. The third-generation detectors like Einstein Telescope could probe such signals at cosmological distances. Direct GW detections in future and indirect EM modeling of GRBs will play complementary roles, and will be the key to advancing our understanding of magnetar physics.

\section*{Acknowledgement}
We acknowledge the use of the public data from the Neil Gehrels Swift data archive, and the UK Swift Science Data Center at the University of Leicester. We appreciate the anonymous referee's valuable comments and suggestions that allow us to improve the manuscript. We also thank Dr. Shuang Du very much for the valuable discussion and help of modifying the manuscript. This work was supported by the National Natural Science Foundation of China (Nos. U1938116, 12173031, 12494572, and 12221003), the Natural Science Foundation of Henan Province (No. 242300420261 and 252300421999), and key scientific research projects of colleges and universities in Henan Province (No. 24a160001), the Fundamental Research Funds for the Central Universities (No. 20720240152), and the Fund of National Key Laboratory of Plasma Physics (No. 6142A04240201). FK Peng also acknowledge support from the University Annual Scientific Research Plan of Anhui Province 2023 (2023AH050146), and the Excellent Teacher Training Program of Anhui Province 2023 (YQZD2023007).

\section*{Data Availability}
The data in this work is publicly available.


\bsp	
\label{lastpage}
\end{document}